\documentclass[%
 aip,
 amsmath,amssymb,
 reprint,%
twocolumn%
] {revtex4-1}

\usepackage{graphicx}% Include figure files
\usepackage{dcolumn}% Align table columns on decimal point
\usepackage{bm}% bold math
\usepackage{xcolor} 
\usepackage[T1]{fontenc}

\begin{document}

\preprint{J. Appl. Phys.}

\title[Critical current density in V-Ti-Y alloys]{Enhancement of functional properties of V$_{0.6}$Ti$_{0.4}$ alloy superconductor by the addition of yttrium}
% Force line breaks with \\

\author{SK. Ramjan}
 \affiliation{Free Electron Laser Utilization Laboratory, Raja Ramanna Centre for Advanced Technology, Indore - 452 013, India.}
 \affiliation{Homi Bhabha National Institute, Training School Complex, Anushakti Nagar, Mumbai 400 094, India.}%
 \author{L. S. Sharath Chandra}%
\email{lsschandra@rrcat.gov.in}
\affiliation{Free Electron Laser Utilization Laboratory, Raja Ramanna Centre for Advanced Technology, Indore - 452 013, India.}%
\author{Rashmi Singh}
\affiliation{Nano-Functional Materials Laboratory, Laser and Functional Materials Division, Raja Ramanna Centre for Advanced Technology, Indore 452 013, India}%
\author{P. Ganesh}
\affiliation{Materials Engineering Laboratory, Raja Ramanna Centre for Advanced Technology, Indore - 452 013, India.}%
\author{Archna Sagdeo}
\affiliation{Hard X-ray Applications Laboratory, Synchrotrons Utilization Section, Raja Ramanna Centre for Advanced Technology, Indore-452 013, India}%
\affiliation{Homi Bhabha National Institute, Training School Complex, Anushakti Nagar, Mumbai 400 094, India.}%
\author{M. K. Chattopadhyay}
\affiliation{Free Electron Laser Utilization Laboratory, Raja Ramanna Centre for Advanced Technology, Indore - 452 013, India.}%
\affiliation{Homi Bhabha National Institute, Training School Complex, Anushakti Nagar, Mumbai 400 094, India.}%

\date{\today}% It is always \today, today,

\begin{abstract}
We show here that the yttrium is immiscible and precipitates with various sizes in the body centred cubic V$_{0.6}$Ti$_{0.4}$ alloy superconductor. The number and size of the precipitates are found to depend on the amount of yttrium added. Precipitates with various sizes up to 30~$\mu$m are found in the V$_{0.6}$Ti$_{0.4}$ alloy containing 5 at.\% yttrium. The large amount of line disorders generated by the addition of yttrium in this alloy are found to be effective in pinning the magnetic flux lines. While the superconducting transition temperature increases with the increasing amount of yttrium in the V$_{0.6}$Ti$_{0.4}$ alloy, the critical current density is maximum for the alloy containing 2 at. \% yttrium, where it is more than 7.5 times the parent alloy in fields higher than 1~T. We found that the effectiveness of each type of defect in pinning the flux lines is dependent on the temperature and the applied magnetic filed. 
\end{abstract}

\maketitle

\section{Introduction}
The V$_{1-x}$Ti$_{x}$ alloys are promising materials as an alternate to the Nb based superconductors for high field applications \cite{tai2007superconducting, matin2015critical, matin2013magnetic, paul2021grain, paul21} especially in the neutron radiation environment. \cite{ivanov1992structural, sekula1978effect, higashiguchi1985microstructure, weber1982neutron, audi2003nubase} Moreover, the V$_{1-x}$Ti$_{x}$ alloys are highly machinable and ductile. \cite{tai2007superconducting, takeuchi2008multifilamentary, bellin1970critical} However, the critical current density (\textit{{J}$_{C}$}) of the V$_{1-x}$Ti$_{x}$ alloys is about \( 10^{7} A/m^{2} \) at 4~K, which is two orders of magnitude less than the commercially available Nb-based superconductors. \cite{matin2015critical} Previous attempts to increase the $J_C$ of these alloys by the introduction of defects through the addition of transition and non-transition elements were ineffective. \cite{efimov1970superconducting} Recently, we have shown that the poly-crystalline V$_{1-x}$Ti$_{x}$ alloys form with large grains having sizes ranging from few $\mu$m to few millimetres. \cite{matin2015critical, matin2013magnetic} We have established that the low grain boundary density and the presence of flux flow channels in the V$_{1-x}$Ti$_{x}$  alloys are the main reasons for the low \textit{{J}$_{C}$} of these alloys. \cite{matin2015critical, matin2013magnetic}

The rare-earth elements are found to be immiscible in vanadium and titanium. \cite{Immiscible, komjathy, yre, smi88,  bus77, cha10, collin59, peng2017formation} Buschow showed by estimating the energy of formation that no binary compound containing rare earth and titanium or vanadium will form. \cite{bus77}  The solubility of rare earths in the liquid vanadium or titanium is limited to very low concentrations (< 1 \%). \cite{yre, smi88, cha10, collin59} We have used this property to introduce a large number of pinning centres by adding gadolinium in the V$_{0.6}$Ti$_{0.4}$ alloy, which resulted in the enhancement of the \textit{{J}$_{C}$} by about 20 times. \cite{paul2021grain} However, the gadolinium precipitates order ferromagnetically \cite{paul21}, which seem to hinder the attempts to improve the $J_C$ further.

\par In this direction, here we present a detailed study on the yttrium (non-magnetic) containing V$_{0.6}$Ti$_{0.4}$ alloys and establish a correlation between the microstructure and the physical properties in the normal and superconducting states. We show that there is an enhancement of \textit{{J}$_{C}$} of V$_{0.6}$Ti$_{0.4}$ by about  7.5 times in fields higher than 1~T when 2 at.\% yttrium is added. Our analysis shows how different defects contribute to the enhancement of \textit{{J}$_{C}$} of the (V$_{0.6}$Ti$_{0.4}$)-Y alloys at different temperatures and field regimes. 

\begin{table}[h]
	\caption{Sample name and the at.\% of elements in the samples}
	\begin{tabular}{p{2cm}p{2cm}p{2cm}p{1cm}}
		\hline\\
		Name of the Sample  & \multicolumn{3}{c}{at.\% of elements} \\
		\cline{2-4}\\ & V & Ti & Y  \\
		\hline\\
		Y0 & 60 & 40 & - \\
		Y1 & 59 & 40 & 1 \\
		Y2 & 58 & 40 & 2\\
		Y3 & 57 & 40 & 3 \\
		Y5& 55 & 40 & 5 \\
		\hline
	\end{tabular}
	\label{tb1}
\end{table}

\section{Experimental}

A series of  samples were synthesized by arc melting \cite{paul2021grain, matin2013magnetic, matin2014influence} the constituent elements (purity better than 99.9\%) in 99.999\% pure Ar atmosphere. The elemental compositions (in at.\%) of the samples are presented in table \ref{tb1}. The samples were cut using diamond saw. The details of metallography experiments can be found elsewhere. \cite{ matin2015critical, matin2015high} Images of the etched samples were taken using a high power optical microscope (Leica DMI 5000M). The elemental analysis was done using the energy dispersive analysis of x-rays (EDAX) setup attached to the scanning electron microscope (SEM, Carl Zeiss, Germany). X-Ray diffraction (XRD) measurements were performed using $\lambda$ = 0.817~\AA~ radiation from the BL12 beamline of the Indus-2 synchrotron facility \cite{sin00}. The resistivity and heat capacity of the samples were measured using the 9~T Physical Property Measurement System (PPMS, Quantum Design, USA). For resistivity measurements, the sample sizes were about 8~mm $\times$ 1~mm $\times$ 0.5~mm. The magnetization was measured using Superconducting Quantum Interference Device based Vibrating Sample Magnetometer (MPMS-3 SQUID-VSM, Quantum Design, USA). The samples used for heat capacity and magnetization measurements have sizes about 3~mm $\times$ 1~mm $\times$ 1~mm.

\begin{figure}
	\centering
	\includegraphics[scale=.18]{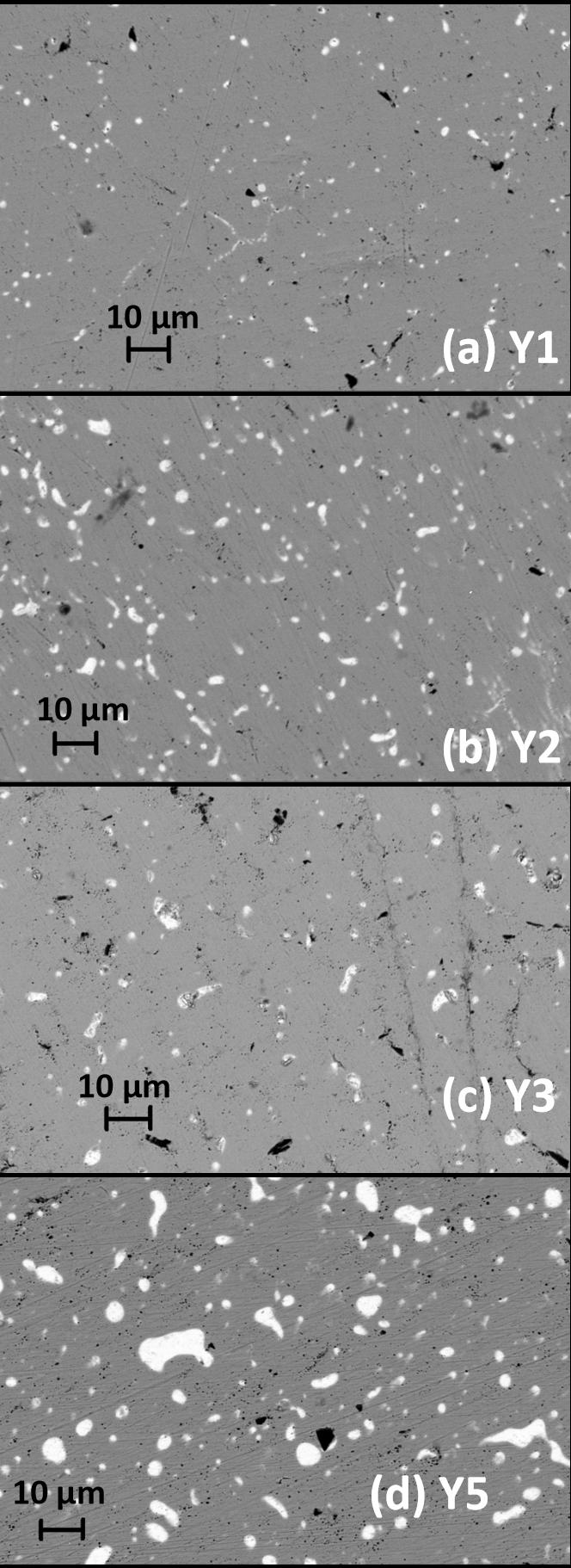}
	\caption{(a-d) SEM images of the polished surfaces of yttrium containing alloys before etching. The size of precipitates (white patches) in the V$_{0.6}$Ti$_{0.4}$ alloy increases with the increasing yttrium content. }
	\label{SEM}
\end{figure}

\begin{figure}
	\centering
	\includegraphics[scale=.35]{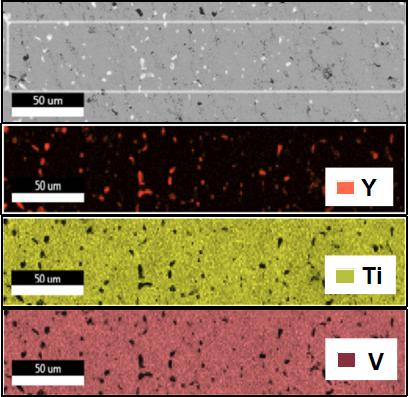}
	\caption{Elemental analysis of the Y3 sample showing immiscibility of yttrium in the V$_{0.6}$Ti$_{0.4}$ alloy.}
	\label{EA}
\end{figure}

\section{Results and Discussion}

\subsection{Microstructural characterization}

Figure \ref{SEM} shows the SEM images of the (V$_{0.6}$Ti$_{0.4}$)-Y alloys up to 5 at.\% of yttrium. These images were taken before etching the samples. Precipitation of secondary phase (white patches) were observed in all the yttrium containing samples. Studies by Love \cite{yre} as well as by Komjathy and coworkers \cite{komjathy} on the vanadium-rare earth and titanium-rare earth binaries reveal that the rare earth elements precipitate in  the vanadium or titanium matrix. The elemental analysis of a portion of Y3 is shown in Fig. \ref{EA}. The white precipitates in Fig.\ref{SEM} are rich in yttrium. The titanium and vanadium are uniformly distributed as a matrix. From Fig.\ref{SEM}, we see that the precipitation follows certain pattern in the Y1 and Y2 samples. This pattern is partially lost in the Y3 sample, whereas the precipitates are more uniformly distributed in the Y5 sample.  The size of the precipitates are presented in table \ref{tb2}. While, the precipitate size in the Y1 and Y2 samples are more or less uniform, there is a distribution of precipitate sizes in the Y3 and Y5 samples. Few precipitates in Y5 are found to be bigger than 30~$\mu$m in size. 

\begin{figure}
	\centering
	\includegraphics[scale=0.6]{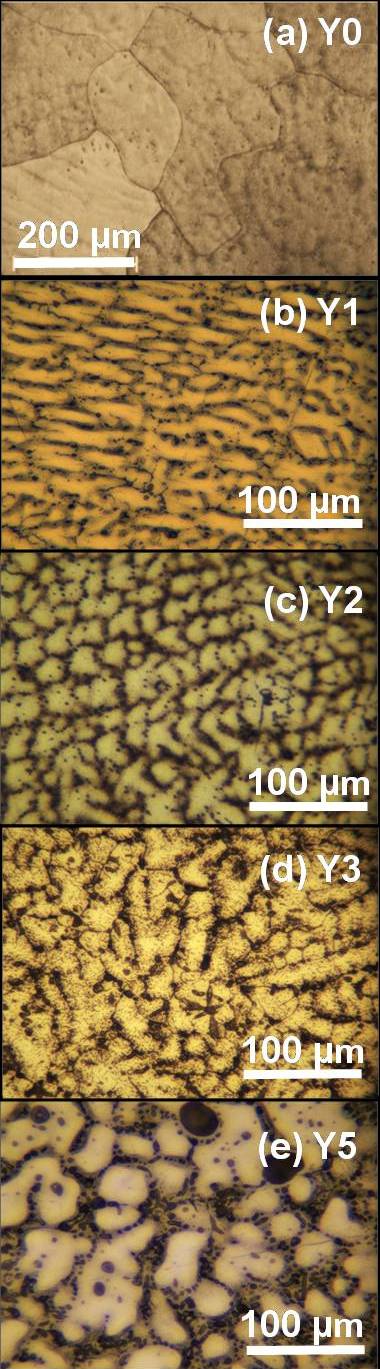}
	\caption{(a-e) Optical metallography images of the polished (V$_{0.6}$Ti$_{0.4}$)-Y alloys after etching. Dendritic growth in alloys containing yttrium indicates the presence of large amount of disorders. The average dendritic cell  size reduces initially with increasing yttrium content, but increases for 3 at.\% yttrium or higher.}
	\label{OM}
\end{figure}

\begin{figure}
	\centering
	\includegraphics[scale=.33]{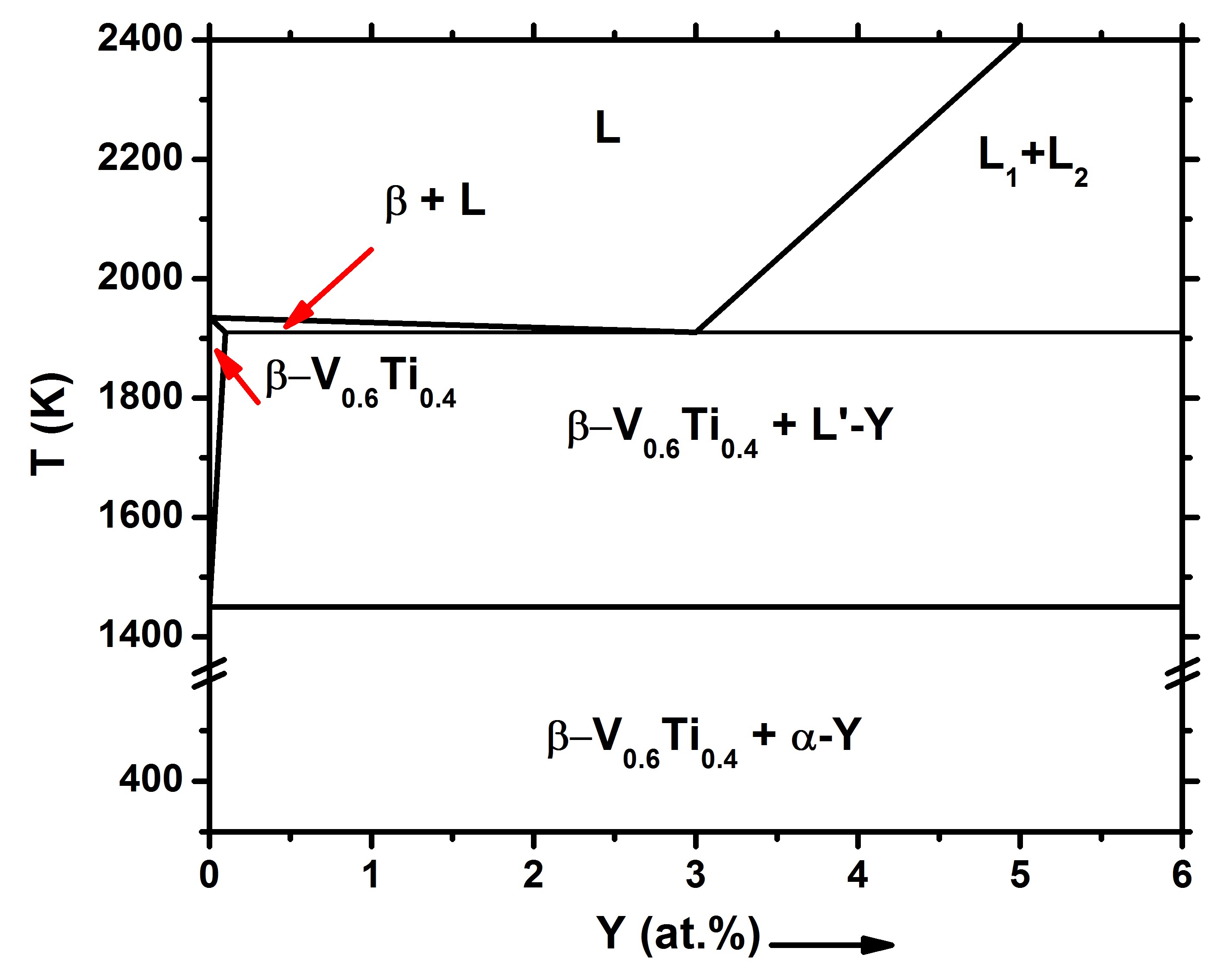}
	\caption{schematics of the phase diagram of the pseudo binary dilute (V$_{0.6}$Ti$_{0.4}$)-Y alloys. At low yttrium content, the homogenous liquid formed at high temperatures phase-separates into a yttrium-rich liquid and solid V$_{0.6}$Ti$_{0.4}$ when cooled below 1935~K. This results in the formation of fine yttrium-rich precipitates below 1430~K.}
	\label{PD}
\end{figure}

The optical metallography images in Fig. \ref{OM} of the (V$_{0.6}$Ti$_{0.4}$)-Y alloys reveal the microstructure.  These images were taken after etching the polished sample surfaces by a solution containing water, HF and HNO$_3$ in the 98:1:1 volume ratio. Dendritic growth is observed in all the alloys containing yttrium indicating a spacial compositional variation. The grain size of the V$_{0.6}$Ti$_{0.4}$ alloy is about 200-300~$\mu$m. The dendritic cell size in Y1, Y2, Y3 and Y5 are 31$\pm$9, 24$\pm$7, 32$\pm$10 and 43$\pm$13 microns respectively. Further studies are required to establish whether these cells are different grains. Nevertheless, thin layers of an alloy which has the lowest melting point, solidifies in between these cells. The cell size reduces initially with increasing yttrium content. This indicates that these yttrium-rich precipitate hinders the cell growth due to the presence of strain field. Thus, there is an enhancement of cell boundary density in the (V$_{0.6}$Ti$_{0.4}$)-Y alloys up to 2 at.\% yttrium addition and yttrium is precipitated only along the cell boundaries. The cell size is found to increase with further addition of yttrium.  The precipitates in these alloys are of larger size and this reduces the number density of precipitates and the strain field in these alloy compositions.

\begin{table}
	\caption{Average sizes of the precipitates }
	\begin{tabular}{p{2cm}p{2.5cm}}
		\hline
		Sample & Precipitate size  \\
		& ($\mu$m)  \\
		\hline\\
		Y1 & 1.3$\pm$0.6   \\
		Y2 & 2.3$\pm$1  \\
		Y3 & 2.8$\pm$1.8  \\
		Y5 & 4.7$\pm$3.3  \\
		\hline
	\end{tabular}
	\label{tb2}
\end{table}

 The vanadium alloys containing small amount of yttrium ($<$ 0.5 at.\%) undergo monotectic transition at about $T_{mt}$ = 2173~K (1900~$^0$C) from homogeneous liquid above $T_{mt}$ to solid $\beta$-vanadium phase and a yttrium-rich liquid below $T_{mt}$. \cite{cha10} The enrichment of yttrium in this liquid leads to phase separation of yttrium-rich and vanadium-rich liquids. If the enrichment of yttrium in the yttrium-rich liquid reaches 96 at.\%, an eutectic transition at about 1730~K ($\approx$ 1460~$^0$C) occurs within this phase. \cite{cha10} On the other hand, the titanium alloys containing small amount of yttrium ($<$ 20 at.\%) undergoes eutectic transition at about $T_{et}$ = 1673~K (1400~$^0$C). \cite{sav62} The melting point ($T_m$) of yttrium is about 1793~K (1520~$^0$C). \cite{cha10} 

The V$_{0.6}$Ti$_{0.4}$ alloy has a $T_m$ of about 1933~K (1660~$^0$C) and the $\beta$-phase (body centred cubic) is stabilized below $T_m$. \cite{melting point} Our metallography results match closely with those of the V-Y alloys indicating that the phase diagram of  the (V$_{0.6}$Ti$_{0.4}$)-Y alloys must be similar to the V-Y phase diagram. On comparing the  literature available on the vanadium/titanium-rare earth binary phase diagrams \cite{yre, komjathy} and our experimental observations, we present a schematic phase diagram for the dilute (V$_{0.6}$Ti$_{0.4}$)-Y alloys in Fig. \ref{PD}.

\begin{figure}[htb]
	\centering
	\includegraphics[width=7cm,height=8cm]{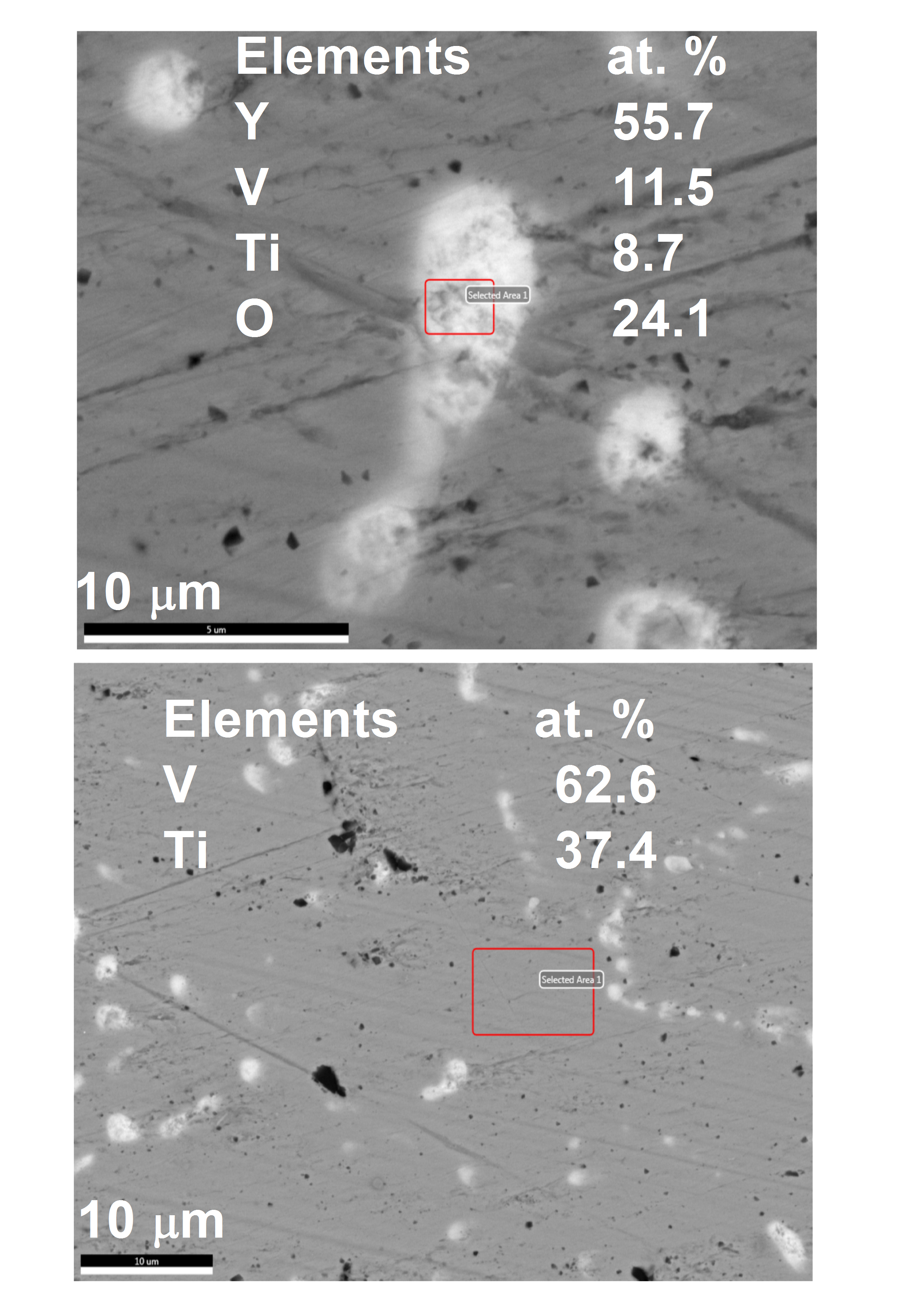}
	\caption{Compositional analysis of yttrium-rich precipitates and V$_{0.60}$Ti$_{0.40}$ matrix using EDAX. Presence of oxygen is found in many of the yttrium-rich precipitates. No trace of yttrium is found in the matrix.}
	\label{ELCMP}
\end{figure}

\begin{figure}[htb]
	\centering
	\includegraphics[width=8cm]{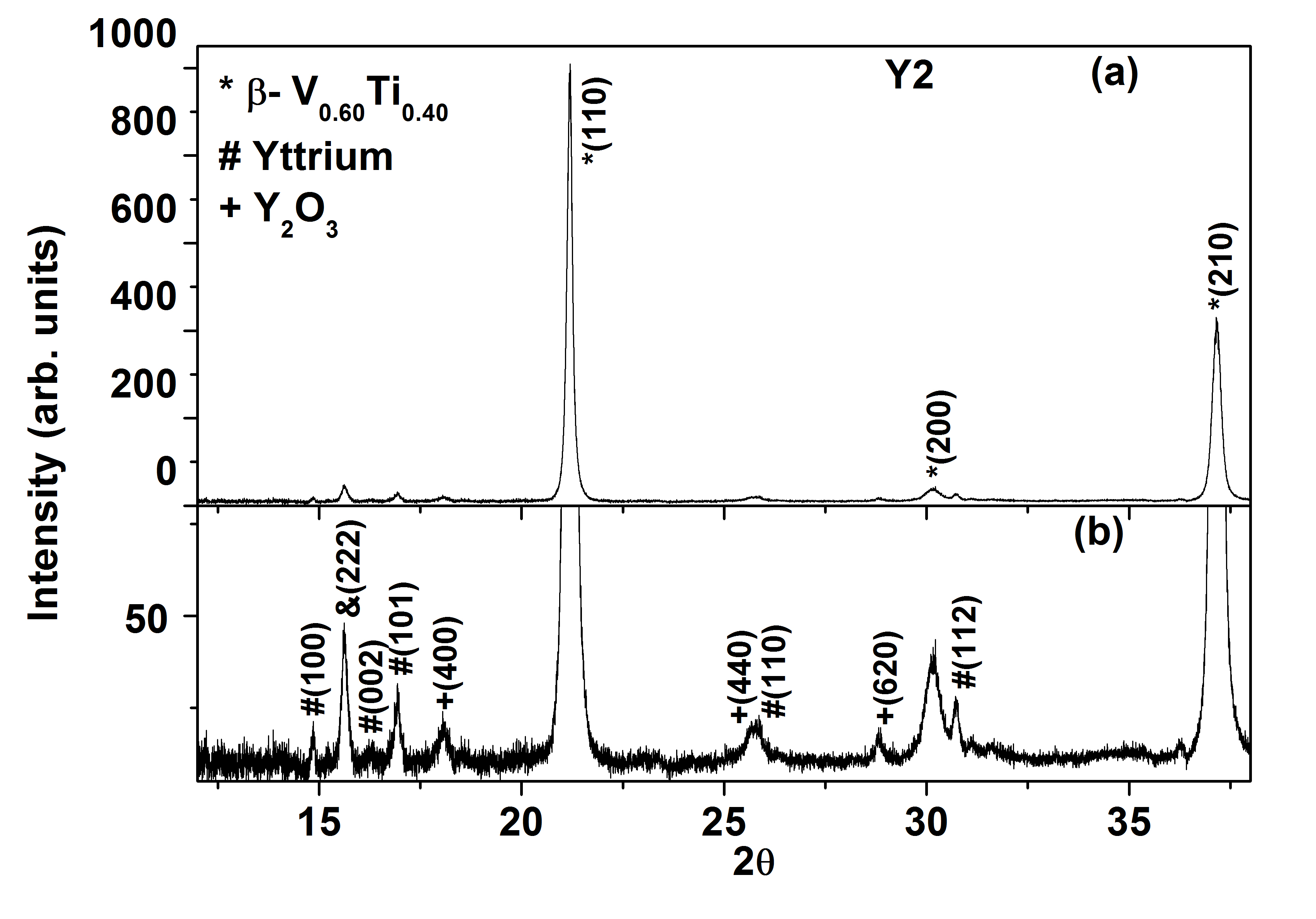}
	\caption{X-ray diffraction pattern of the Y2 alloy (a) The major peaks are indexed to $\beta$- V$_{0.60}$Ti$_{0.40}$ phase. (b) Weak reflections corresponding to $\alpha$-Y and Y$_2$O$_3$ phases are also seen.}
	\label{XRD}
\end{figure}

The small size of the precipitates in the alloys containing 3 at.\% or less yttrium indicates that these alloys, while cooling from the melt undergo monotectic transition from the homogeneous V-Ti-Y liquid to solid $\beta$-V-Ti alloy and a yttrium-rich liquid. Figure \ref{ELCMP} shows the typical compositions of the precipitates and matrix of the Y2 alloy.  The composition of the matrix is about V$_{0.626}$Ti$_{0.374}$ with a minor local variation of the composition. The oxygen content within a precipitate varies substantially among the precipitates. We found that higher the oxygen content, higher is the amount of titanium and vanadium present in the precipitate. In the cases where oxygen is absent in the precipitates, the composition is close to 95 at. \% of yttrium. This indicates that eutectic microstructure may be present within the yttrium-rich precipitates. \cite{cha10} In order to get a clear picture of the different phases present in the samples, we show in Fig. \ref{XRD}, the x-ray diffraction (XRD) pattern of the Y2 alloy.  The symbols '$\star$', '$\#$' and '+' represent reflections from the $\beta$-V$_{0.60}$Ti$_{0.40}$, yttrium precipitates and Y$_2$O$_3$ phases respectively. The lattice parameter of $\beta$-V$_{0.60}$Ti$_{0.40}$ is about 3.1412 \AA~while that of Y$_2$O$_3$ is about 10.4 \AA. The lattice parameters of the yttrium precipitates are about $a$ = 3.647 \AA~and $c$ = 5.728 \AA. The lattice parameters of $\beta$-V$_{0.60}$Ti$_{0.40}$ and yttrium precipitates are in agreement with literature.  \cite{spe56, paul21} On the other hand, the lattice parameter of Y$_2$O$_3$ is slightly less than that of bulk which may be due to oxygen off-stiochiometry. \cite{han84} Thus, the reduction of the cell size in these compositions is related to the formation of fine yttrium-rich precipitates during the phase separation of the homogeneous liquid into solid $\beta$-V$_{0.6}$Ti$_{0.4}$ and L$'$-Y phases (Fig. \ref{PD}). On the other hand, a two phase microstructure seen in between the large V-Ti cells in the Y5 sample (Fig.\ref{OM}.(e)) is caused by the liquid immiscibility.  

Earlier, we have shown that the V$_{1-x}$Ti$_x$ alloys have large grain sizes of the the order of few microns to few millimetres. Our microstructural studies reveal that the yttrium can introduce large amount of defects in the V$_{1-x}$Ti$_x$ alloys which is helpful in improving the critical current density in the mixed state of these superconducting alloys. Therefore, we characterize  the present alloys in the superconducting as well as normal states.

\subsection{Electrical and thermal properties of the (V$_{0.6}$Ti$_{0.4}$)-Y alloys and the superconducting transition temperature ($T_C$)}

\begin{figure}[htb]
	\centering
	\includegraphics[width=8cm]{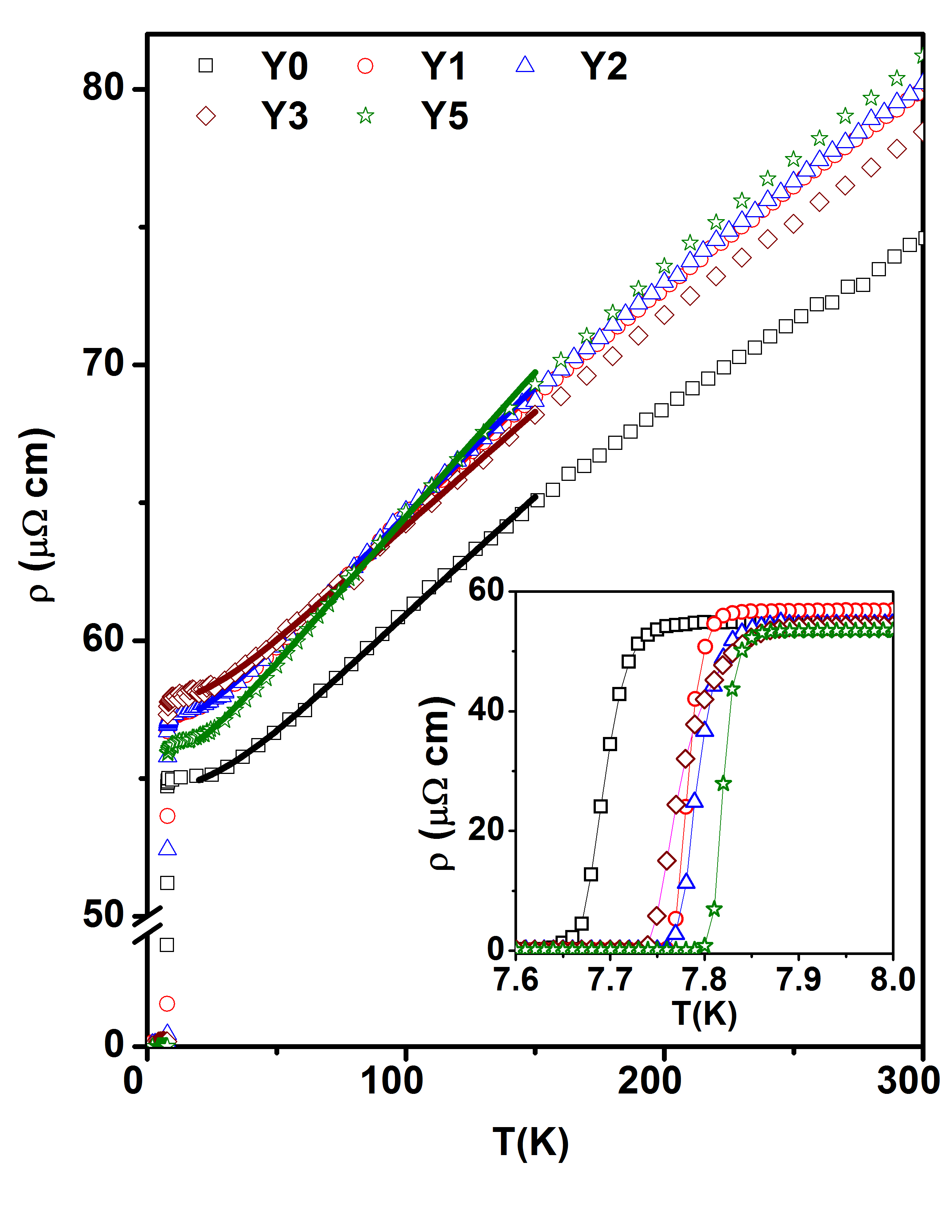}
	\caption{Temperature dependence of electrical resistivity of the (V$_{0.6}$Ti$_{0.4}$)-Y alloys. The residual resistivity and critical temperature increases with yttrium addition. Open symbols are the experimental data points and the solid lines are the fits using eq.(1). The parameters of fitting are presented in table \ref{tb3}. The inset shows expanded view of the resistivity around the superconducting transition.}
	\label{RT}
\end{figure}

Figure \ref{RT} shows the temperature dependence of electrical resistivity ($\rho$($T$)) of the (V$_{0.6}$Ti$_{0.4}$)-Y alloys in the range 2-300~K. The yttrium containing alloys have higher $\rho$($T$) in comparison with that of the parent V$_{0.6}$Ti$_{0.4}$ alloy. Residual resistivity ($\rho_0$) increases up to 3 at.\% yttrium due to the increased static defects (precipitates, grain/cell boundaries, dislocations  and point defects). The reduced number density of defects in the Y5 alloy due to the larger size of the precipitates results in the lower $\rho_0$ as compared to the other yttrium containing alloys. The inset to Fig. \ref{RT} shows the expanded view of resistivity around the \textit{T}$_{C}$. The \textit{T}$_{C}$ is obtained as that temperature at which the temperature derivative of resistivity shows a peak. The \textit{T}$_{C}$ increases from 7.68~K for the V$_{0.6}$Ti$_{0.4}$ alloy to 7.85~K for the Y5 sample. 

Normally, the $T_C$ is expected to decrease with increasing disorder. \cite{aoki1967non, aoki1969non} However, the increase in the $T_C$ in the present alloys can be due to  (i) suppression of spin fluctuations by the disorder, \cite{matin2014influence, stritzker1979superconductivity,bose1990effect} (ii) increase in the electron-phonon coupling due to phonon softening by the defects, \cite{shy15} and/or (iii) removal of the trace oxygen by yttrium from the V-Ti matrix. \cite{peng2017formation,collin59}

The \textit{T}$_{C}$ of the V$_{1-y}$Ti$_{y}$ alloys is limited by the spin fluctuations \cite{matin2014influence} and increases from 5.4~K for $y$ = 0 to about 7.68~K for the $y$ = 0.4 alloy. We have argued that the suppression of spin fluctuation by the disorder introduced when vanadium is alloyed with titanium increases the $T_C$ of the V$_{1-y}$Ti$_{y}$ alloys. \cite{matin2014influence} Palladium is also found to be superconducting when certain type of disorder is introduced to suppress the spin fluctuations. \cite{stritzker1979superconductivity,bose1990effect} Thus, the increase in the \textit{T}$_{C}$ of the yttrium containing alloys can be attributed to the suppression of spin fluctuations with increase in disorder. However, this should result in the reduction of the normal state $\rho$($T$), \cite{paul21} which is contrary to the observed result that the yttrium containing alloys have higher normal state $\rho$($T$) as compared to the parent V$_{0.6}$Ti$_{0.4}$ alloy. The normal state $\rho$($T$) in the range 20-120~K is fitted using the equation\cite{rice67, shy15}:

\begin{eqnarray}
\rho(T) = \rho_0 &+& \rho_{sf}\left(\frac{T}{T_{sf}}\right)^2F_{2}\left(\frac{T_{sf}}{T}\right)\nonumber \\
&-&\rho_{sf}\left(\frac{T}{T_{sf}}\right)^5F_{5}\left(\frac{T_{sf}}{T}\right)\nonumber \\
&+&\rho_{sd}\left(\frac{T}{\theta_{D}}\right)^3F_{3}\left(\frac{\theta_{D}}{T}\right)
\end{eqnarray}

where $\rho_0$ is residual resistivity, $\rho_{sf}$ is the coefficient of resistivity for spin-fluctuation, $\rho_{sd}$ is the coefficient of resistivity corresponding to interband scattering, $T_{sf}$ is the spin fluctuation temperature, $\theta_D$ is the Debye temperature and $F_{k}$($M$) is the Fermi-integral given by 

\begin{equation}
F_{k}(M) = \int_{0}^{M} \frac{{\text d} z z^{k}\exp(z)}{(\exp(z)-1)^2}
\end{equation}  

The fitting is shown in Fig. \ref{RT} and coefficients obtained by the fitting are presented in the table \ref{tb3}. The errors in the estimation of parameters are within 10\%. We found that $\rho_{sf}$ increases with increasing yttrium content. The large $\theta_D$ and small $\rho_{sd}$ for Y3 and Y5 alloys indicate a negligible contribution from the interband scattering to resistivity in these alloys. The $T_{sf}$ also increases with yttrium content up to 3 at. \%. Therefore, the enhancement of $T_C$ in the yttrium containing alloys is not due to the suppression of spin fluctuations.

\begin{table}[h]
	\caption{Parameters obtained from the fitting of resistivity}
	\begin{tabular}{|c|c|c|c|c|c|}
		\hline
		Sample & $\rho_0$ & $\rho_{sf}$ & $\rho_{sd}$ & $T_{sf}$ & $\theta_{D}$ \\
					& ($\mu \Omega $ cm) & ($\mu \Omega$ cm) & ($\mu \Omega$ cm)& (K) & (K) \\ 
		\hline
		Y0 & 54.6 & 9.1 & 25.8&196&300 \\
		Y1 & 56.9 & 18.0 & 14.0&198&366 \\
		Y2 & 56.9 & 19.6 & 12.3&204&382\\
		Y3 & 57.7 & 30.4 & 3.4&308&561 \\
		Y5& 55.7 & 26.1 & 7.9&217&433 \\
		\hline
	\end{tabular}
	\label{tb3}
\end{table}

\begin{figure}[htb]
	\centering
	\includegraphics[width=8cm]{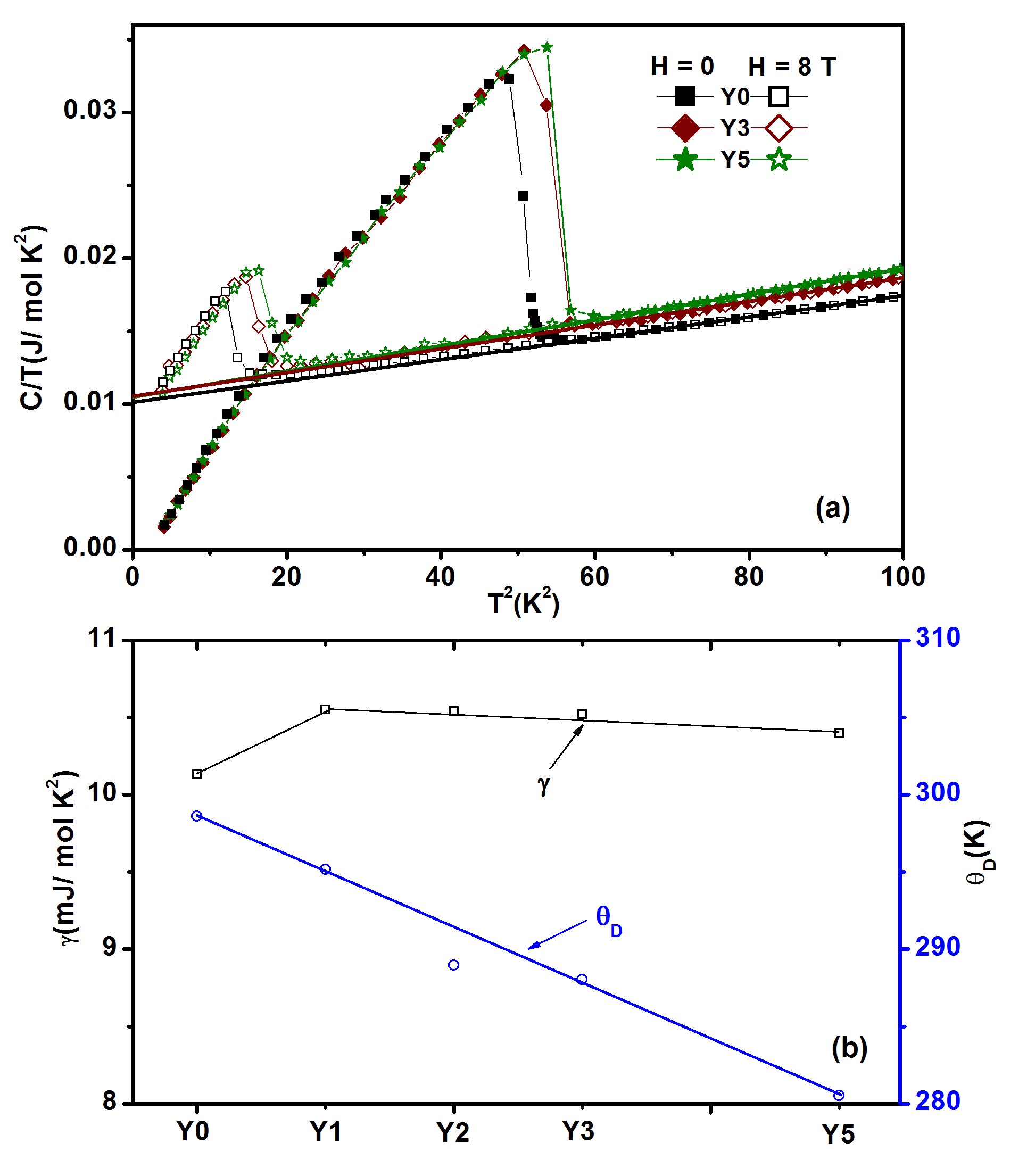}
	\caption{(a) Temperature dependence of heat capacity of the V$_{0.6}$Ti$_{0.4}$, Y3 and Y5 alloys measured in the zero and 8~T magnetic fields. The symbols are the experimental data points and the solid lines are the linear fits. (b) Sommerfeld coefficient $\gamma$ and Debye temperature $\theta_D$ as a function of yttrium content in the present alloys. The solid lines are guide to eye.}
	\label{ct}
\end{figure}

The higher $\rho$($T$) with a different curvature for the yttrium containing alloys as compared to that of the parent V$_{0.6}$Ti$_{0.4}$ alloy hints at the enhancement of electron-phonon coupling due to the softening of phonon modes by disorder. To verify the correlation between the $T_C$ and phonon softening, we have measured the heat capacity of the yttrium containing alloys in zero and 8~T fields (Fig. \ref{ct}(a)). The normal state $C/T$ v/s $T^2$ data in 8~T in the range 5-10~K is used to fit a straight line to obtain the Sommerfeld coefficient of electronic heat capacity ($\gamma$) and the Debye temperature ($\theta_D$) using the relation $ C = \gamma T + \beta T^3$ where $\theta_D^3 = 1943.66/\beta$. \cite{matin2014influence} The maximum change in the $\theta_D$, $\gamma$ (Fig. \ref{ct}(b)) and $T_C$ with the yttrium content are about 6.5\%,  4\%, and 2.3\% respectively. Since, change in $T_C$ is quite smaller in comparison with the other two variables, it is unlikely that the enhancement of the $T_C$  is due to changes in the electron-phonon coupling by the addition of yttrium.  

The $T_C$ of vanadium is about 5.4~K. \cite{tcv} The presence of oxygen suppresses the $T_C$ of vanadium. \cite{wexler1952superconductivity} It is well known that the addition of 0.5-2 at.\% yttrium in titanium or vanadium improves the ductility. \cite{collin59} This improvement is caused by the scavenger effect of yttrium in removing oxygen from the grain boundaries of vanadium and titanium. \cite{collin59} Therefore, we infer that the addition of yttrium to the V$_{1-x}$Ti$_x$ alloys improves the $T_C$ by removing oxygen from the matrix. Our studies on the (V$_{0.6}$Ti$_{0.4}$)$_{50}$Y$_{50}$ alloy showed that the superconductivity is induced in yttrium-rich phase by the proximity effect \cite{ram21hfpme}. Therefore, we infer that the region of yttrium-rich precipitates where the oxygen is absent become superconducting below $T_{C}$ due to proximity effect and the boundaries between the precipitates and the matrix can act as effective pinning centres.    

%ooooooooooooooooooooooooooooooooooooooooooooooooooooooooooooo
\subsection{Magnetic properties of the (V$_{0.6}$Ti$_{0.4}$)-Y alloys: Role of microstructure on the enhancement of critical current density}

\begin{figure}[htb]
	\centering
	\includegraphics[width=8cm]{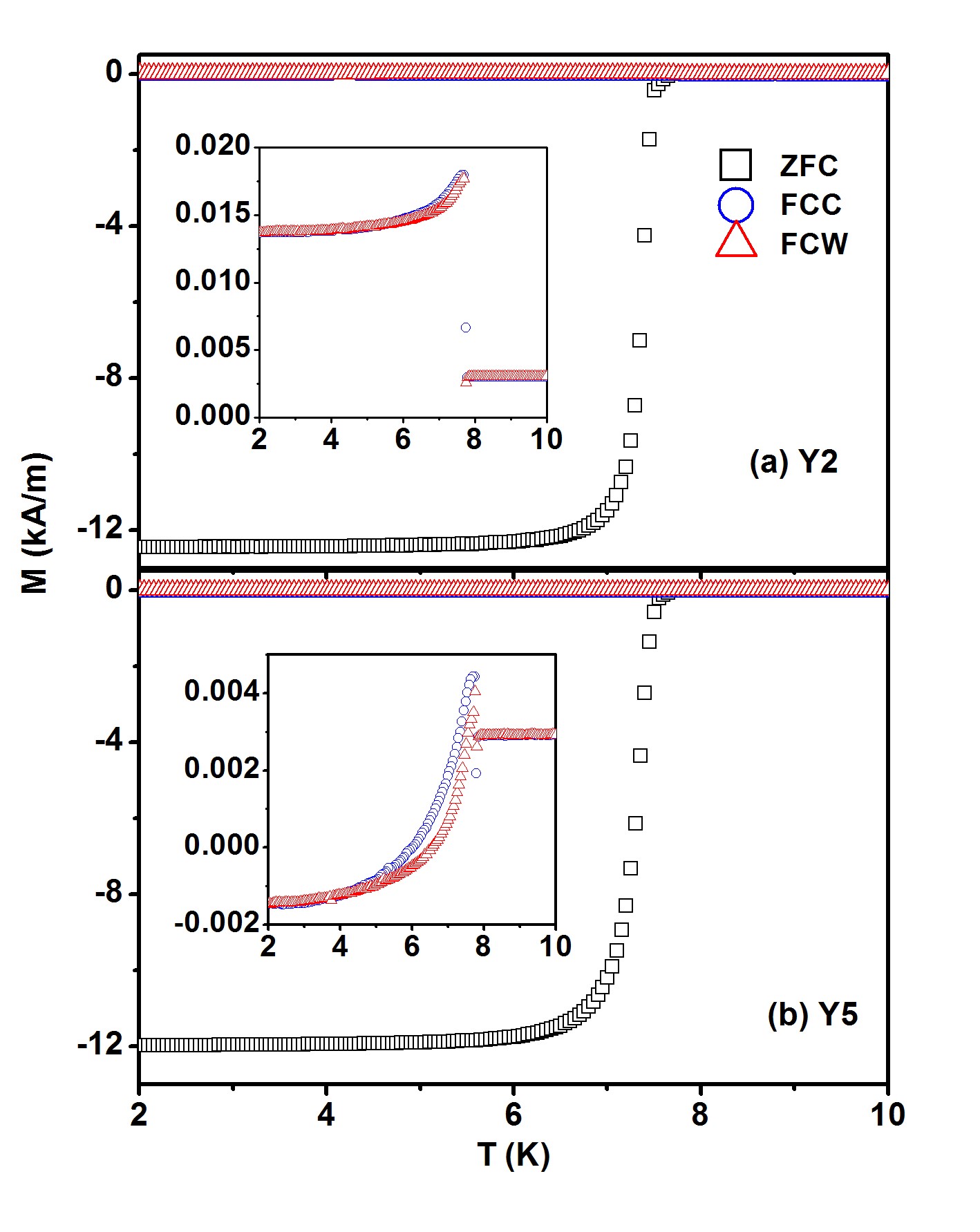}
	\caption{Temperature dependence of magnetisation of (a) Y2 and (b) Y5 in the temperature range 2-10 K measured in the presence of 10~mT field. The insets show the expanded view of the FCC and FCW response in 10~mT field.}
	\label{MT}
\end{figure}

\begin{figure}[h]  
	\includegraphics[width=8cm]{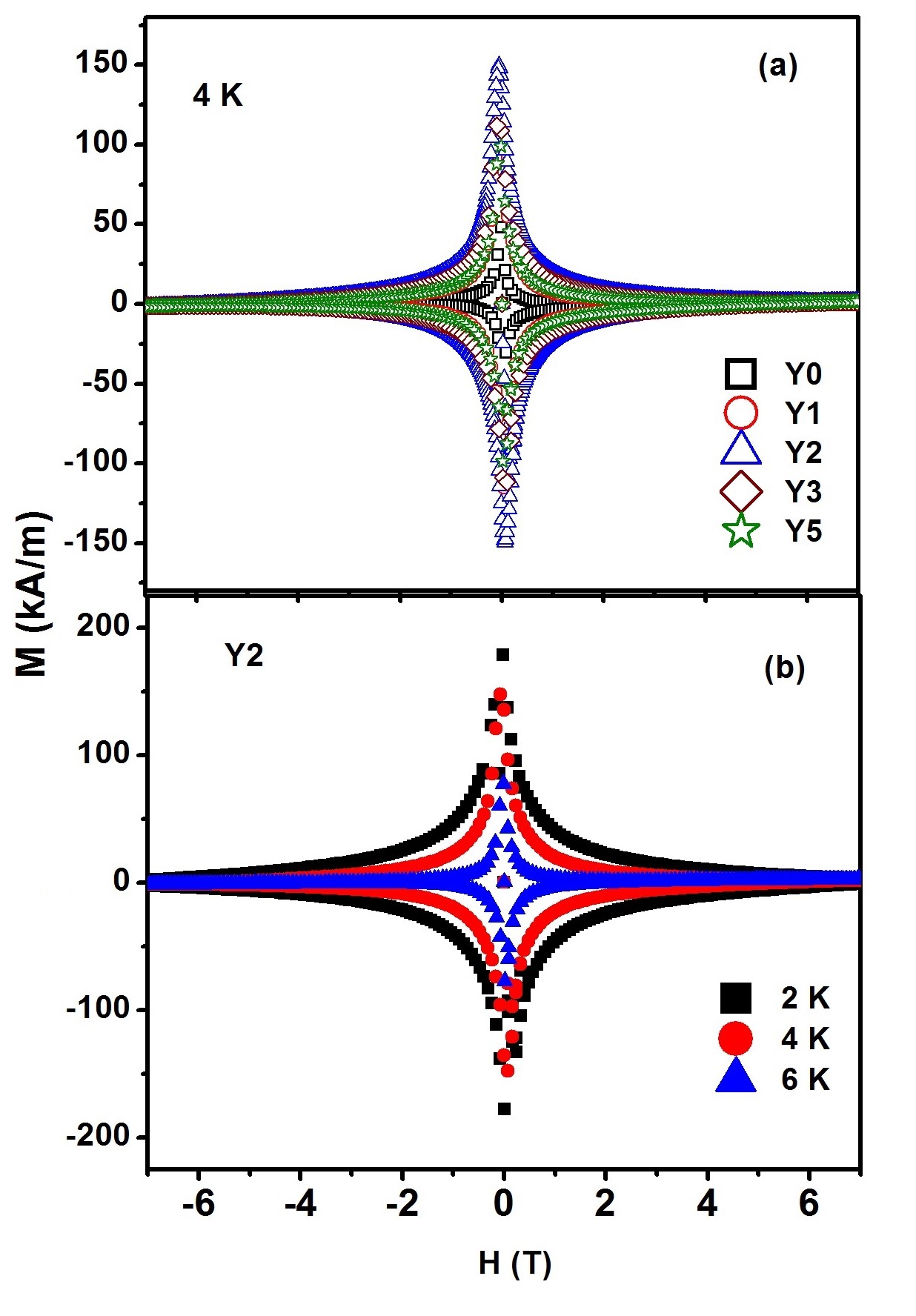}
	\caption{(a) Magnetic field dependence of magnetisation for the (V$_{0.6}$Ti$_{0.4}$)-Y alloys at 4~K. The hysteresis is observed to increase with yttrium addition and is maximum for the Y2 alloy. (b) Magnetic field dependence of magnetisation for  the Y2 alloy at 2~K, 4~K and 6~K.}
	\label{MH}
\end{figure} 

The temperature dependence of magnetization ($M$($T$)) of the Y2 and Y5 alloys are shown in Fig. \ref{MT}. The $M$($T$) is measured in the presence of 10~mT field while warming up after cooling down the sample from $T > T_C$ to 2~K in zero field (ZFC), while cooling down in the same field after warming above $T_C$ (FCC), and then again while warming up in the same field (FCW). The \textit{T}$_{C}$ is estimated as that temperature at which the $M$($T$) starts to decrease towards negative values when the temperature is decreased from 10~K. The \textit{T}$_{C}$ estimated from  $M$($T$) is in agreement with that estimated from the resistivity measurements. The insets to the Fig. \ref{MT} show the expanded view of $M$($T$) measured during FCC and FCW cycles. The Meissner fraction ($M_f$) is estimated as the ratio $M_{FCC}$/$M_{ZFC}$ at 2~K. The $M_f$ is about 0.095\% for the V$_{0.6}$Ti$_{0.4}$ alloy which decreases with increasing amount of yttrium in this alloy. The $M_{FCC}$/$M_{ZFC}$ is about 0.032\% for the Y3 alloy indicating that the magnetic flux line pinning improves when yttrium is added to the V$_{0.6}$Ti$_{0.4}$ alloy. This indicates that the yttrium containing alloys have higher $J_C$ than the parent V$_{0.6}$Ti$_{0.4}$ alloy.    

In order to quantify the the enhancement of $J_C$ in the yttrium containing alloys, we have measured the field dependence of magnetisation ($M$($H$)) for all the alloys at different temperatures. The Fig. \ref{MH}(a) shows the $M$($H$) at 4~K for all the alloys. The size of the hysteresis increases with increasing yttrium content in the V$_{0.6}$Ti$_{0.4}$ alloy up to 2  at.\%, and then it starts shrinking. Figure \ref{MH}(b) shows the $M$($H$) of the Y2 alloy at different temperatures. The hysteresis in the $M$($H$) is symmetric along the $H$ axis indicating that the Bean-Livingston surface barrier effect \cite{bean1964surface} is negligible in Y2 alloy. The upper critical field ($H_{C2}$) and magnetic irreversibility field ($H_{irr}$) at various temperatures below $T_C$ are estimated from the isothermal magnetisation curves. The magnetic field at which $M$($H$) deviates from its behaviour in the normal state is taken as the $H_{C2}$. The magnetic field at which the $M$($H$) for increasing $H$ bifurcates from that during the $H$ decreasing cycle is taken as the $H_{irr}$.

\begin{figure}
	\includegraphics[width=8cm]{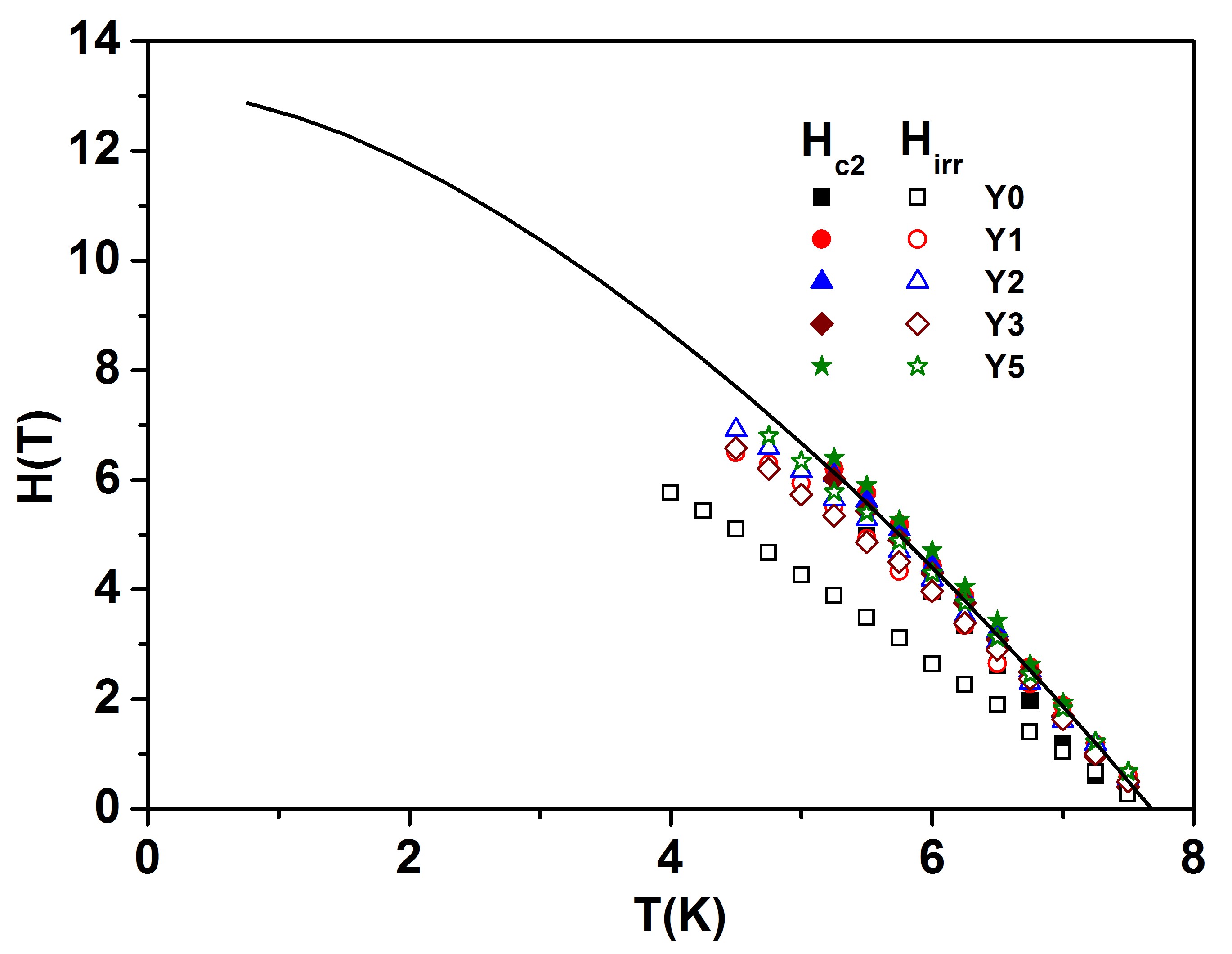}
	\caption{The temperature dependence of upper critical fields (closed symbols) and irreversibility fields (open symbols) for the (V$_{0.6}$Ti$_{0.4}$)-Y alloys. The solid line is the fit to the $H_{C2}$ using WHH formalism.}
	\label{hc2}
\end{figure}

The $H_{C2}$ and $H_{irr}$ as a function of temperature for all the alloys are plotted in Fig. \ref{hc2}. The $H_{C2}$($T$) is almost the same for all the yttrium containing alloys and $H_{irr} < H_{C2}$. 
Addition of yttrium to V$_{0.6}$Ti$_{0.4}$ alloy increases the $H_{irr}$. The solid line in Fig. \ref{hc2} represents the fit to the $H_{C2}$($T$) using the Werthamer-Helfand-Hohenberg (WHH) formalism for dirty limit superconductors. \cite{werthamer1966temperature} From the fitting, the parameters $\alpha$ (corresponding to the Pauli paramagnetic effect) and $\lambda_{SO}$ (which is the measure of strength of spin-orbit interaction) are found to 1.49 and 2.5 respectively. The \textit{H}$_{C2}$ in the limit of absolute zero (\textit{H}$_{C2}(0)$) is found to be about 13.2~T which is comparable to the Nb-Ti alloys. \cite{poo00}  

%oooooooooooooooooooooooooooooooooooooooooooooooooooooooooooo

\begin{figure}[h]
	\includegraphics[width=8cm]{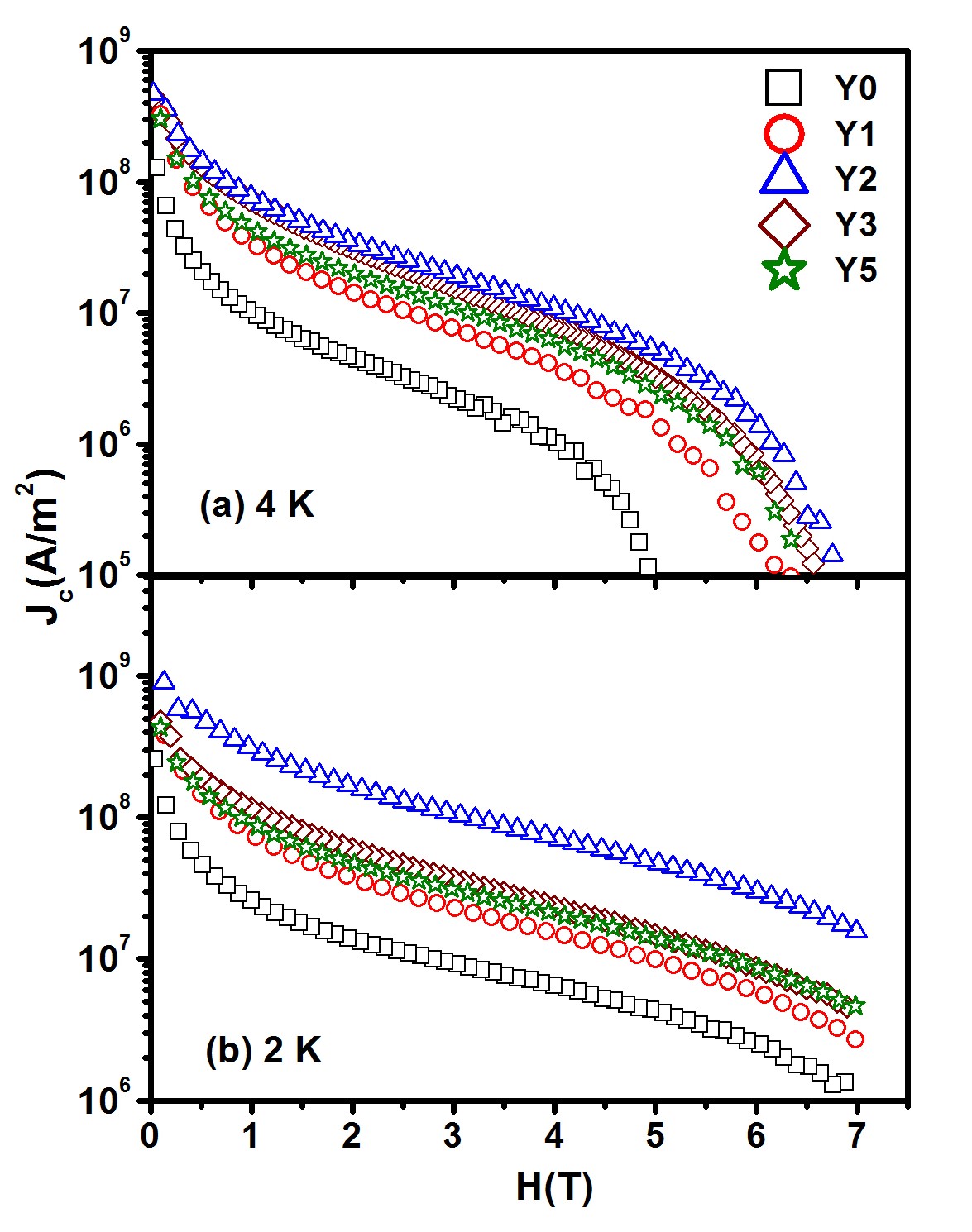}
	\caption{Field dependence of critical current density of the (V$_{0.6}$Ti$_{0.4}$)-Y alloys at (a) 4~K and (b) 2~K. The $J_C$ increases by approximately 8 times at 2~K when 2~at.\% yttrium is added to V$_{0.6}$Ti$_{0.4}$.}
	\label{CCD}
\end{figure}

The critical current density (\textit{{J}$_{C}$}) of the (V$_{0.6}$Ti$_{0.4}$)-Y alloys estimated from the $M$($H$) is shown in Fig.\ref{CCD}. The \textit{{J}$_{C}$} is estimated using the Bean's critical state model as \cite{zheng1995reversible,martinez2007flux,sundar2015magnetic,bean1964magnetization} 

\begin{equation}
	J_{C}=2\Delta M \left[a\left(1-\frac{a}{3b}\right)\right]^{-1}.
\end{equation}

Here, the $\Delta M$ at every $H$, is the difference between the $M$ measured during increasing and decreasing $H$ cycles.  The parameters \textit{a} and \textit{b} ($b>a$) are the dimensions of the rectangular cross section of the sample in the direction normal to the applied magnetic field.

The zero field \textit{{J}$_{C}$} value at 2~K for the V$_{0.6}$Ti$_{0.4}$ alloy is estimated to be $ 2\times10^8$~A/m$^2$, which is in close agreement with the literature. \cite{matin2015critical} The \textit{{J}$_{C}$} increases significantly with the addition of yttrium up to 2 at.\%, and then it decreases with further addition. The Y2 alloy has a \textit{J$_{C}$} of $ 7\times10^6~A/m^2 $ at 4~K and 1~T field, which is an order of magnitude higher than that of the V$_{0.6}$Ti$_{0.4}$ alloy.  The drop in \textit{{J}$_{C}$} with the application of low magnetic fields is found to be less steep for all the yttrium containing alloys in comparison with the parent V$_{0.6}$Ti$_{0.4}$ alloy. The $J_C$ of the parent V$_{0.6}$Ti$_{0.4}$ alloy exists only up to 5~T at 4~K. On the other hand, the yttrium containing alloys have significant $J_C$ above 5~T at 4~K. The value of $J_C$ as well as the range over which a significant $J_C$ is present increase with the increasing yttrium content up to 2~at.\%. The $J_C$ at 2~K (Fig.\ref{CCD}) exceeds 2$\times$10$^7$~Am$^{-2}$ at 7~T for the Y2 alloy. The increase in the defects when yttrium is added to the V$_{0.6}$Ti$_{0.4}$ alloy (section 3.1) results in the enhancement of flux line pinning,  which in turn increases the $J_C$. Though the enhancement of $J_C$ by adding yttrium in V$_{0.6}$Ti$_{0.4}$ alloy is significant, the $J_C$ is still considerably lower than that of commercial Nb-Ti alloys \cite{sha19, end73, che91, bou06, miy06, lin15, mou17}.    

\begin{figure}
	\centering
	\includegraphics[width=8cm]{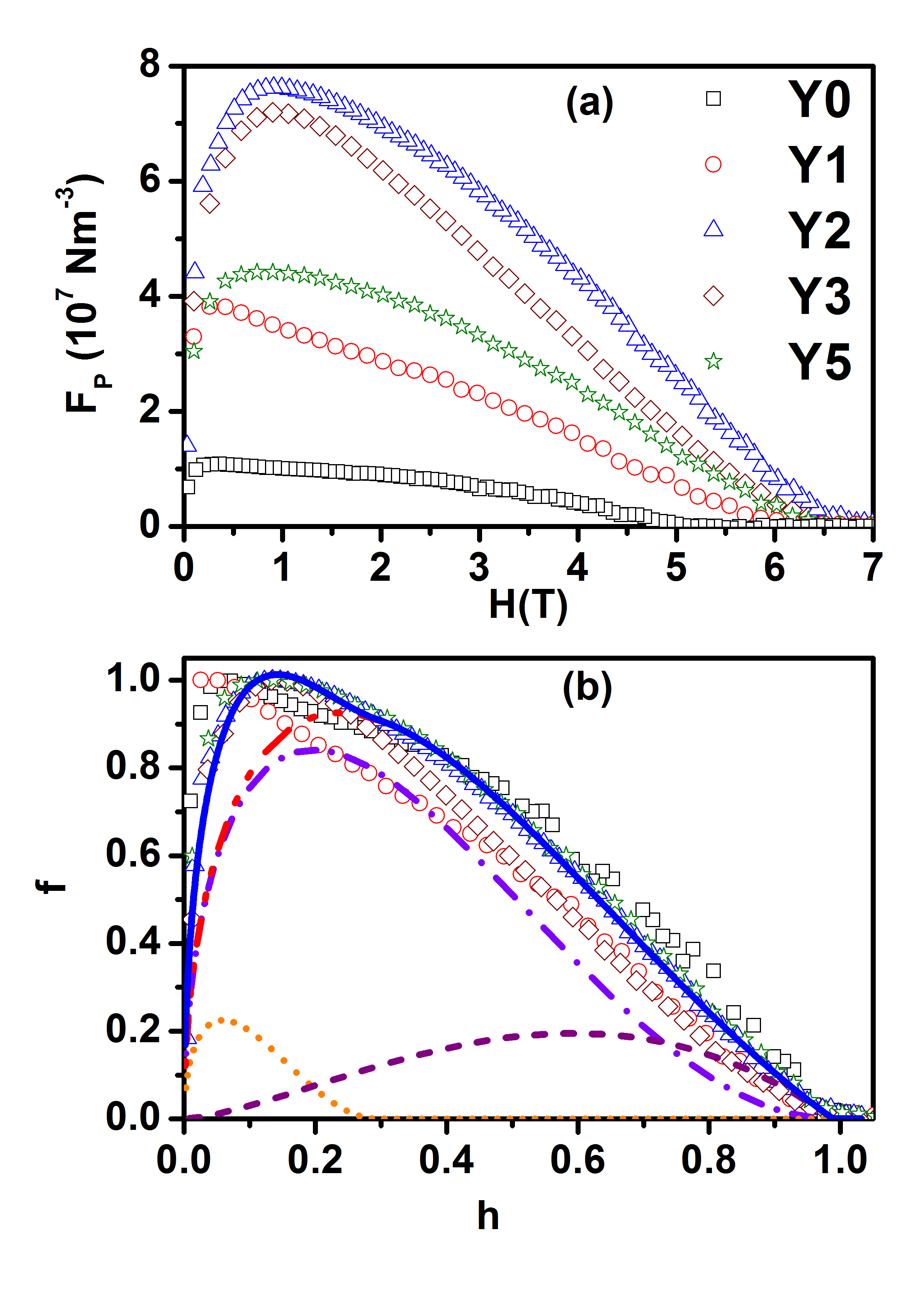}
	\caption{ (a) Field dependence of pinning force density for the (V$_{0.6}$Ti$_{0.4}$)-Y alloys at 4~K. Addition of yttrium results in the enhancement of the maximum pinning force density by about 8 times. (b) The reduced pinning force density as a function of reduced field from the plots in (a). }
	\label{FP}
\end{figure}

\begin{figure}[htb]
	\includegraphics[width=8cm]{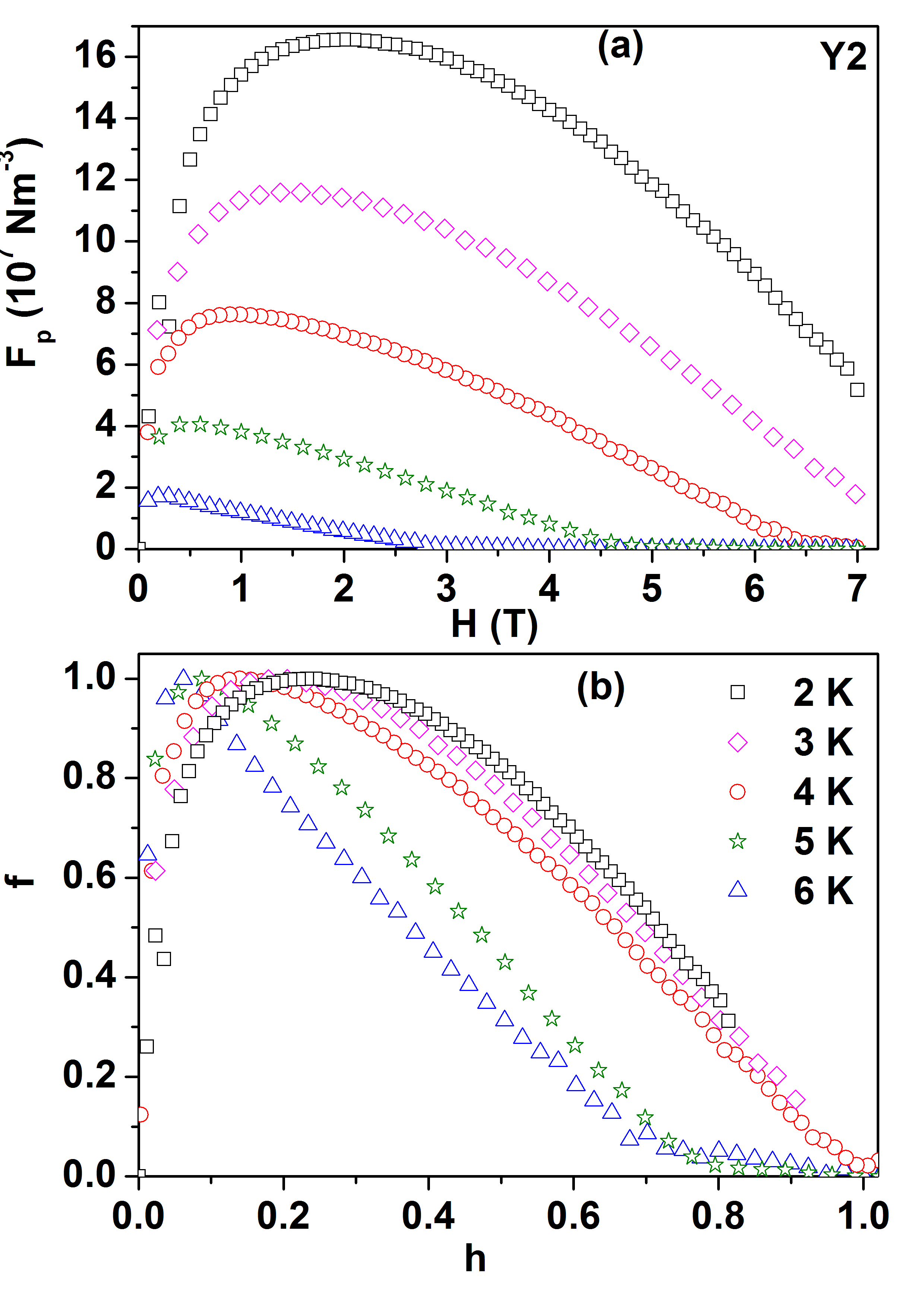}
	\caption{(a) Field dependence of $F_p$ of Y2 at different temperatures. The shape of the $F_p$ curves changes with temperature. (b) The reduced pinning force density as a function of reduced field from the plots in (a). Large improvement in the pinning force density at high magnetic fields is achieved below 5~K.}
	\label{FPA}
\end{figure}

In order to establish a correlation between the nature of the defects and the enhancement of $J_C$, we have estimated the pinning force density ($F_P$) at 4~K as $F_P = J_C \times H$ and the same is plotted in Fig. \ref{FP}(a). In the V$_{1-y}$Ti$_y$ alloys, the grain boundaries are the major pinning centres in the low-field regime, whereas the point defects and dislocations are the effective pinning centres in the high-field regime. \cite{matin2015critical,matin2013magnetic} The maximum $F_P$ of the V$_{0.6}$Ti$_{0.4}$ alloy at 4~K is in the range of 10$^7$ Nm$^{-3}$. Addition of yttrium increases this value up to 2 at.\%. The maximum $F_P$ of the Y2 alloy at 4~K is about 7.6$\times$10$^{7}$~Nm$^{-3}$ which is about 7.6 times that of the parent V$_{0.6}$Ti$_{0.4}$ alloy. Significant pinning strength above 5~T is found in the yttrium containing alloys. In all the samples at 4~K, the $F_p$ increases sharply in very low fields and falls of gradually in high magnetic fields.

The Fig. \ref{FP}(b) presents the plot of reduced pinning force density  ($f=F_{p}/F_{p-max}$) of all the alloys as a function of reduced field $h = H/H_{irr}$. According to Dew-Hughes \cite{dewhughes}, the $f$ depends on the spacing, size and nature of the pinning centres and is proportional to $h^{p}(1-h)^{q}$. Each type of pinning centre yields an unique set of \textit{p} and \textit{q} values \cite{ekin2010unified} and a maximum of $f$ at $h_m = p/(p+q)$. The values of $p$, $q$ and $h_m$ provides information on the nature of the pinning centres responsible for the $J_C$. All the curves in Fig. \ref{FP}(b) tend to scale on the falling edge, while significant variation of $f$ among the alloys is observed for $h <$ 0.1. This indicates that the pinning mechanism at high magnetic fields is same for all the alloys. The experimental value of $h$ at which $F_p$ becomes maximum ($h_m$) for the V$_{0.6}$Ti$_{0.4}$ alloy is about 0.06.  The $h_m$ increases with the addition of yttrium and is about 0.14 for the Y2, Y3 and Y5 alloys at 4~K. The comparison of the curvature of the different pinning mechanisms for $h$ close to unity \cite{dewhughes} with the curvature of $f$ of the present alloys indicates that the pinning at high fields is due to the regions with large change in the superconducting properties (large change in the Ginsburg-Landau parameter $\kappa$ or large $\Delta\kappa$). \cite{dewhughes, ekin2010unified} The known mechanisms of flux pinning have $h_m \geq$ 0.2. The $h_m < $ 0.2 for all the alloys indicates that the pinning that exists at low magnetic fields may not be effective at high fields. 

In such cases where  multiple types of pinning centres become available for pinning at different magnetic fields, $F_p$ can be expressed as,\cite{muz11, li06}
 \begin{equation}
F_{p}  = F_{p1} +F_{p2} +F_{p3} +.......
 \end{equation}
where, $F_{pi}$ is the pinning force density of individual pinning centres. The solid blue line in Fig. \ref{FP}(b)  shows the fit to the $f$ by considering three types of pinning centres viz., \\
(i)  normal surface pinning ($h^{0.5}(1-h)^{2}$) by grain/cell boundaries (violet dot-dashed line)\\
(ii) $\Delta \kappa$ surface pinning ($h^{1.5}(1-h)$) by dislocations (purple dashed line) and\\
(iii) normal surface pinning with $H'_{irr} = 0.22 \times H_{irr}$ (orange dotted line).\\

One can see that pinning centres of type (i) and (ii) combined (red dash-dot-dotted line in Fig. \ref{FP}(b)) can account for pinning in the high field range $h >$ 0.35. The variation of $F_{p-max}$ with the composition follows the cell boundary and dislocation density in these alloys. The $F_{p}$ is maximum for the Y2 alloy which has the smallest cell size. The number of precipitates is largest for Y2 alloy which hints that the dislocation density may be maximum for this alloy. Thus, the grain/cell boundaries and dislocations are the effective pinning centres in all the present alloys.  Therefore, the $f$ versus $h$ plots for all the alloys at 4~K scale at high fields. To explain the pinning in the low field region, one needs to consider pinning mechanism of type (iii) with a very low irreversibility field $H'_{irr}$ which is about 0.22$ \times H_{irr}$. We have observed a positive magnetization signal just below $T_{C}$($H$) of all the (V$_{0.6}$Ti$_{0.4}$)-Y alloys in low magnetic fields. \cite{ram21hfpme} This is an effect what is known as 'High Field Paramagnetic Effect (HFPME)' where the positive magnetization arises from the flux compression as well as creeping of flux lines from weak pinning-centres to strong pinning-centres leading to anisotropic distribution in flux line pinning. \cite{matinHFPME, shyamHFPME, Dias} The flux compression is found to be due to the yttrium-rich precipitates become superconducting by the proximity effect. \cite{ram21hfpme} We believe that the pinning mechanism of type (iii) is related to the pinning centres that contributes to HFPME. 

The shape of the $F_p$ curve and its peak position change with temperature (Fig. \ref{FPA}(a)). The position of the maximum $F_p$ changes from $h_m$ = 0.08 at 6~K to about 0.23 at 2~K (Fig. \ref{FPA}(b)). The $F_p$ curves are extrapolated linearly to get $h_m$ ($H_{irr}$) at temperature 3~K and 2~K. The sharpness of the peak reduces and the curvature of the falling edge changes with decreasing temperature. This behaviour is an indication of different pinning centres becoming effective at different temperatures. The Fig. \ref{FPA}(b) shows that the pinning mechanism present in high fields and low temperatures is absent at temperatures close to $T_{C}$. The thermal conductivity studies on (V$_{0.6}$Ti$_{0.4}$)-Gd alloys indicated that the dislocations become effective in scattering the phonons at low temperatures and high magnetic fields which in turn renormalizes the electron-phonon coupling. \cite{paul21} We infer that the dislocation become effective in pinning flux lines in high magnetic fields at low temperatures which is inline with the analysis related to Fig. \ref{FP}(b). 

%ooooooooooooooooooooooooooooooooooooo
\section{Conclusion}
We have shown that the yttrium is immiscible and precipitates with various sizes in the V$_{0.60}$Ti$_{0.40}$ alloy. Dendritic microstructure is observed in all the yttrium containing alloys. For $\leq$ 2 at.\% yttrium in the V$_{0.60}$Ti$_{0.40}$ alloy, fine yttrium-rich precipitates are generated because of phase separation of the homogeneous V-Ti-Y liquid into a solid $\beta$-V$_{0.60}$Ti$_{0.40}$ alloy and a yttrium-rich liquid. The size of the yttrium-rich precipitates increases for higher yttrium content due to liquid immiscibility. The dendritic cell size reduces with increasing yttrium content up to 2 at.\% in the V$_{0.60}$Ti$_{0.40}$ alloy which results in the generation of a large number of line defects. Yttrium removes oxygen from the V$_{0.60}$Ti$_{0.40}$ alloy matrix due to which the $T_C$ of the yttrium containing alloys is enhanced. The defects generated by the addition of yttrium are found to be effective in pinning the flux lines and also increase the $H_{irr}$. The critical current density is increased by more than 7.5 times in fields higher than 1~T for the V$_{0.58}$Ti$_{0.40}$Y$_{0.02}$ alloy at 4~K.

%\vskip -1 cm
%\nocite{*}
%\bibliography{aipsamp}% Produces the bibliography via BibTeX.

\section{References}
\begin{thebibliography} {}

\bibitem{tai2007superconducting} M. Tai, K. Inoue, A. Kikuchi, T. Takeuchi, T. Kiyoshi, Y. Hishinuma, IEEE. Trans. Appl. Supercond. \textbf{17}, 2542 (2007).

\bibitem{matin2015critical} Md. Matin, L. S. Sharath Chandra, M. K. Chattopadhyay, R. K. Meena, R. Kaul, M. N. Singh, A. K. Sinha and S. B. Roy, Physica C \textbf{512}, 32 (2015).

\bibitem{matin2013magnetic} Md. Matin, L. S. Sharath Chandra, M. K. Chattopadhyay, R. K. Meena, R. Kaul, M. N. Singh, A. K. Sinha and S. B. Roy, J. Appl. Phys. \textbf{113}, 163903 (2013).

\bibitem{paul2021grain} S. Paul, SK. Ramjan, R. Venkatesh, L. S. Sharath Chandra and M. K. Chattopadhyay,  IEEE. Trans. Appl. Supercond. \textbf{31}, 8000104 (2021). 

\bibitem{paul21} S. Paul, SK. Ramjan, L. S. Sharath Chandra, M. K. Chattopadhyay, Mater. Sci. Eng. B (accepted) [arXiv preprint arXiv:2103.13601 (2021)].

\bibitem{ivanov1992structural} L. I. Ivanov, V. V. Ivanov, V. M. Lazorenko, Y. M. Platov and V. I. Tovtin, J. Nucl. Mater. \textbf{191}, 928 (1992).

\bibitem{sekula1978effect} S. T. Sekula, J. Nucl. Mate. \textbf{72} 91 (1978) 

\bibitem{higashiguchi1985microstructure} Y. Higashiguchi, H. Kayano and S. Morozumi, J. Nucl. Mater. \textbf{133}, 662 (1985).  

\bibitem{weber1982neutron} W. H. Weber, Jour. Nucl. Mate. \textbf{108}, 572 (1982).

\bibitem{audi2003nubase} G. Audi, O. Bersillon, J. Blachot and A. D. Wapstra, Nucl. Phys. A \textbf{624}, 1 (2003).

\bibitem{takeuchi2008multifilamentary} T. Takeuchi, H. Takigawa, M. Nakagawa, N. Banno, K. Inoue, Y. Iijima and A. Kikuchi, Supercond. Sci. Technol. \textbf{21}, 025004 (2008). 

\bibitem{bellin1970critical} P. H. Bellin, H. C. Gatos and V. Sadagopan, J. Appl. Phys. \textbf{41}, 2057 (1970).

\bibitem{efimov1970superconducting} Y. V. Efimov, V. V. Baron and E. M. Savitskii, \textit{Physics and Metallurgy of Superconductors} (Springer, 1970, pp-98-101).

\bibitem{Immiscible} K. A. Gschneidner, Jr., Structural and physical properties of alloys and intermetallic compounds, in: L. Eyring (Ed.), Progress in the science and technology of rare earths, 1, 1964, Pergamon, New York pp. 222-258.

\bibitem{komjathy} A. S. Komjathy, R. H. Read and W. Rostoker, WADD Technical report no. 59-483, Wright-Patterson Airforce base, Ohio (1960). Permanent Online link: https://catalog.hathitrust.org/Record/009230761.

\bibitem{yre} B. Love, WADD Technical report no. 60-74, part I, Wright-Patterson Airforce base, Ohio (1960). Permanent Online link: https://catalog.hathitrust.org/Record/009206837.

\bibitem{smi88} J. F. Smith, K. J. Lee, and D. M. Martin, Binary rare earth-vanadium systems, CALPHAD 12 (1988) 89-96.

\bibitem{bus77} K. H. J. Buschow, Intermetallic compounds of rare-earth and 3d transition metals, Rep. Prog. Phys. 40 (1977) 1179-1256.

\bibitem{cha10} W. Chan, M. C. Gao, O. N. Dogan, and P. King, Thermodynamic assessment of V-rare earth systems, J. Phase Equilib. Diffus. 31 (2010) 425-432.

\bibitem{collin59} J. F. Collins, V. P. Calkins, and J. A. Mc Gurty, Applications of rare earths to ferrous and non-ferrous alloys. United States: N. p., (1959). Web. doi:10.2172/4215576.

\bibitem{peng2017formation} L. Peng, C. Jiang, X. Li, P. Zhou, Y. Li, and X. Lai, J. Alloy. Compd. \textbf{694} 1165 (2017).

\bibitem{matin2014influence} Md. Matin, L. S. Sharath Chandra, S. K. Pandey, M. K. Chattopadhyay and S. B. Roy, Eur. J. Phys. B \textbf{87}, 131 (2014).

\bibitem{matin2015high} Md. Matin, M. K. Chattopadhyay, L. S. Sharath Chandra, and S. B. Roy, Supercond. Sci. Technol. \textbf{29}, 025003 (2015). 

\bibitem{sin00} A. K. Sinha, A. Sagdeo, P. Gupta, A. Kumar, M. N. Sing, R. K. Gupta, S. R. Kane, S. K. Deb, AIP Conf. Proc. {\bf 1349}, 503 (2011).

\bibitem{sav62} E. M. Savitskii and G. S. Burkhanov, J. Less-Common Metals {\bf 4}, 301 (1962). 

\bibitem{melting point} A. Kostov, D. Zivkovic and B. Friedrich, J. Min. Metall. B {\bf 42} 57 (2006). 

\bibitem{aoki1967non} R. Aoki and T. Ohtsuka, J. Phys. Soc. Japan \textbf{23}, 955 (1967).

\bibitem{aoki1969non} R. Aoki and T. Ohtsuka, J. Phys. Soc. Japan \textbf{26}, 651 (1969).

\bibitem{stritzker1979superconductivity} B. Stritzker, Phys. Rev. Lett. \textbf{42}, 1769 (1979).

\bibitem{bose1990effect} S. K. Bose, J. Kudrnovsky, I. I. Mazin and O. K. Andersen, Phys. Rev. B \textbf{41}, 7988 (1990).

\bibitem{shy15} Shyam Sundar, L. S. Sharath Chandra, M. K. Chattopadhyay S. K. Pandey, D. Venkateshwarlu, R. Rawat, V. Ganesan, and S. B. Roy, New J. Phys. {\bf 17}, 053003 (2015).

\bibitem{tcv} S. Paul, L. S. Sharath Chandra, and M. K. Chattopadhyay, J. Phys.: Condens. Matter {\bf 31} 475801 (2019).

\bibitem{wexler1952superconductivity} A. Wexler and S. W. Corak, Phys. Rev. \textbf{85}, 85 (1952).

\bibitem{bean1964surface} C. P. Bean and J. D. Livingston, Phys. Rev. Lett. \textbf{12}, 14 (1964).

\bibitem{werthamer1966temperature} N. R. Werthamer, E. F. Helfand and P. C. Hohenberg, Phys. Rev. \textbf{147}, 295 (1966).

\bibitem{poo00} eg., C. P. Poole, Jr. Handbook of superconductivity (Academic Press, San Diego, (2000)).

\bibitem{zheng1995reversible} D. N. Zheng, H. D. Ramsbottom and D. P. Hampshire, Phys. Rev. B \textbf{52}, 12931 (1995).

\bibitem{martinez2007flux} E. Mart{\'\i}nez, P. Mikheenko, M. Mart{\'\i}nez-L{\'o}pez, A. Mill{\'a}n, A. Bevan, and J. S. Abell, Phys. Rev. B \textbf{75}, 134515 (2007).

\bibitem{sundar2015magnetic} Shyam Sundar, M. K. Chattopadhyay, L. S. Sharath Chandra and S. B. Roy, Physica C \textbf{519}, 13 (2015).

\bibitem{bean1964magnetization} C. P. Bean, Rev. Mod. Phys. \textbf{36}, 31 (1964).

\bibitem{dewhughes} D. Dew-Hughes, Phil. Mag. \textbf{30}, 293 (1974).

\bibitem{ekin2010unified} J. W. Ekin, Supercond. Sci. Technol. \textbf{23}, 083001 (2010).

\bibitem{ram21hfpme} SK. Ramjan, L. S. Sharath Chandra and M. K. Chattopadhyay, arXiv:2110.05921

\bibitem{muz11} L. Muzzi, G. De Marzi, C. F. Zignani, U. B. Vetrella, V. Corato, A. Rufoloni, and A. della Corte, IEEE Trans. Appl. Supercond. {\bf 21}, 3132 (2011).

\bibitem{li06} P. J. Li, Z. H. Wang, A. M. Hu, Z. Bai, L. Qiu, and J Gao, Supercond. Sci. Technol. {\bf 19}, 825 (2006).

\bibitem{matinHFPME} Md. Matin, L. S. Sharath Chandra, M. K. Chattopadhyay, M. N. Singh, A. K. Sinha, and S. B. Roy, Supercond. Sci. Technol. {\bf 26}, 115005 (2013).

\bibitem{shyamHFPME} S. Sundar, M. K. Chattopadhyay, L. S. Sharath Chandra, and S. B. Roy, Supercond. Sci. Technol. {\bf 28}, 075011 (2015).

\bibitem{Dias} F. T. Dias, P. Pureur, P. Rodrigues Jr. and X. Obradors, Phys. Rev. B {\bf 70}, 224519 (2004).



\bibitem{rice67} A. I. Schindler, and M. J. Rice, Phys. Rev. {\bf 164}, 759 (1967). 

\bibitem{spe56} F. H. Spedding, A. H. Daane, and K. W. Herrmann, Acta Cryst. {\bf 9}, 559 (1956).

\bibitem{han84} F. Hanic, M. Hartmanova, G. G. Knab, A. A. Urusovskaya and K. S. Bagdasarov, Acta Cryst. B {\bf 40}, 76 (1984).

	\bibitem{sha19} L. S. Sharath Chandra, Sabyasachi Paul, Ashish Khandewlwal, Vinay Kaushik, Archna Sagdeo, R. Venkatesh, Kranti Kumar, A. Banerjee, M. K. Chattopadhyay, J. Appl. Phys. {\bf 126}, 183905 (2019).

		\bibitem{end73} G. Enderlein, A. Handstein, F. Lange, and P. Verges, Cryogenics {\bf 13}, 426 (1973).

		\bibitem{che91} O. V. Chernyj, G. F. Tikhinskij, G. E. Storozhilov, M. B. Lazareva, L. A. Kornienko, N. F. Andrlevskaya, V. V. Slezov, V. V. Sagalovich, Ya D. Starodubov, and S. I. Savchenko, Supercond. Sci. Technol. {\bf 4}, 318 (1991). 

		\bibitem{bou06} T. Boutboul, S. Le Naour, D. Leroy, L. Oberli, and V. Previtali, IEEE Trans. Appl. Supercond. {\bf 16}, 1184 (2006).  

		\bibitem{miy06} K. Miyashita, H. Sato, M. Arika, and R. Takahashi, Electr. Eng. Jpn. {\bf 156}, 24 (2006).

		\bibitem{lin15} H. Lin, C. Yao, H. Zhang, X. Zhang, Q. Zhang, C. Dong, D. Wang, and Y. Ma, Sci. Rep. {\bf 5}, 11506 (2015). 

		\bibitem{mou17} T. Mousavi, Z. Hong, A. Morrison, A. London, P. S. Grant, C. Grovenor and S. C. Speller, Supercond. Sci. Technol. {\bf 30}, 094001 (2017).



\end{thebibliography}

\section{Data availability statement}

The data that support the findings of this study are available from the corresponding author upon reasonable request.

\end{document}